# COINTEGRATION OF THE DAILY ELECTRIC POWER SYSTEM LOAD AND THE WEATHER

S. Z. Stefanov

*ESO EAD, 5, Veslets Str., 1040 Sofia, Bulgaria*

szstefanov@ndc.bg

This paper makes a thermal predictive analysis of the electric power system security for a day ahead. This predictive analysis is set as a thermal computation of the expected security. This computation is obtained by cointegrating the daily electric power system load and the weather, by finding the daily electric power system thermodynamics and by introducing tests for this thermodynamics. The predictive analysis made shows the electricity consumers' wisdom.

## 1. Introduction

The electric power system (EPS) is affected by weather changes and by the exchanges with other EPS's. The EPS load is unpredictable.

The load of a network-modelled EPS is dynamically unpredictable. Under this model, the load and the weather are cointegrated, according to Ref.1. Fezzi and Bunn[2] have made a cointegration of the daily load and the wholesale price of electricity. Therefore, there is cointegration of the daily EPS load and the weather.

The load of a field-modelled EPS is thermodynamically unpredictable. Modelling the EPS as a field is possible under the internal model principle from system theory. Under this model the EPS is viewed as an open system. Under this model there is evolution of the EPS behaviour. The thermodynamic unpredictability diminishes when predicting rare events, cooperative and competitive phenomena in the EPS. It is such type of events and phenomena that are predicted by the EPS dispatchers.

An intelligent system can be viewed as a dissipative model of the brain dynamics from Ref.3. This intelligent system has a field computation and a field realization in the sense of Ref.4. That is why it is able to make a predictive analysis of the EPS.

Wehenkel and Pavella,[5] Abed and al.[6] have made a predictive analysis of the security of a network-modelled EPS from the data about the latter. These analyses are incomplete. They do not predict the change in the EPS security, caused by the evolution in the EPS behaviour.

The aim of this paper is a thermal predictive analysis of the EPS security for a day ahead. This analysis is sought by means of cointegration of the daily EPS load and the weather.

## 2. Cointegration

The daily EPS load is modelled by one descriptive and two rescriptive models. These models are constructed for a time, whose moments are the calendar days.

The descriptive load model is a regression with indicators for the two load peaks, a distributed lag which represents the load variability, and a flow integrator in the load. The regression with indicators is

$$P^a_t = a_0 + a_1 P_{t-1} + a_2 P_{t-3/2} + a_3 P_{t-7} + a_4 \delta_9 + a_5 \delta_{10} + a_6 \delta_{19} + a_7 \delta_{20} + a_8 \delta_{11} + a_9 \delta_{21} \qquad (1)$$

In (1), $\delta_i$ denotes a single-amplitude impulse at the i-th hour of the day. Here $P_{t-3/2}$ is the load which, in the morning, is equal to the day-before-yesterday afternoon load, and in the afternoon - to the yesterday morning load.

The distributed lag is for the last two regressors in (1). The flow integrator leads to substitution, in (1), of the regressors $P_{t-2}$ and $P_{t-3}$ by $P_{t-3/2}$.

The rescriptive load models are regressions with cointegration links, a distributed lag which represents the load variability, and a flow integrator in the load. These regressions are

$$P^b_t = b_0 + b_1 P_{t-1} + b_2 P_{t-3/2} + b_3 P_{t-7} + b_4 (P_{t-1} - P_{t-3/2}) T_{t-2} + b_5 (P_{t-3/2} - P_{t-7}) T_{t-8} \qquad (2)$$

$$P^c_t = c_0 + c_1 P_{t-1} + c_2 P_{t-3/2} + c_3 P_{t-7} + c_4 T_{t-2} + c_5 T_{t-8} + c_6 (P_{t-1} - P_{t-3/2}) T^2_{t-2} + c_7 (P_{t-1} - P_{t-3/2}) T_{t-2} +$$

$$c_8 (P_{t-3/2} - P_{t-7}) T_{t-8} \qquad (3)$$

The distributed lag is for the last two regressors in (2), respectively – in (3). The flow integrator leads to substitution, in (2) and (3), of the regressors $P_{t-2}$ and $P_{t-3}$ by $P_{t-3/2}$. Here the regressor $P_{t-3/2}$ has the same meaning as in (1).

Model (1) is descriptive, and models (2) and (3) – rescriptive, in the sense of Ref.7. The regression (2) models the normal load behaviour, and the regressions (1) and (3) – the evolution load behaviour.

Predicting the daily load by these three models is predicting by an ensemble of models, according to Ref.8. Under each of these three models, the data are treated sequentially, and not in parallel. The data are treated in this way because the sequential models of daily load forecast give[9] a better forecast than the parallel ones.

The flow integration reduces the data noise.[10] That is why the regressions (1), (2) and (3) give a more accurate load forecast. The flow in the data presents the EPS exchanges in the load models.

The daily EPS load is econometrically modelled by the regressions (1), (2) and (3). This is modelling of the changes in the load and in the environment. These regressions are models of the dynamic unpredictability of the load.

## 3. Daily EPS Thermodynamics

The cointegration distance between the regressions $P^a_t$ and $P^b_t$, respectively $P^c_t$ and $P^b_t$, is set[11] by the angle $\theta_1$, respectively – the angle $\theta_2$,

$$\theta_1 = 0.5 \arctan( 2 <p^a_t p^b_t> / (<p^a_t p^a_t> - <p^b_t p^b_t>)) \qquad (4)$$

$$\theta_2 = 0.5 \arctan( 2 <p^c_t p^b_t> / (<p^c_t p^c_t> - <p^b_t p^b_t>))$$

Here $<\cdot>$ denotes the mean for the time $\tau=1,...,24$, and the time series $p^a_t$, $p^b_t$, $p^c_t$, $\tau=1,...,24$, are obtained from the time series $P^a_t$, $P^b_t$, $P^c_t$, $\tau=1,...,24$, by subtracting the mean.

The EPS entropy S and the environment entropy S´ are[12]

$$S(\chi) = - ((1-\cos\chi)/2)\ln((1-\cos\chi)/2) – ((1+\cos\chi)/2)\ln((1+\cos\chi)/2) \qquad (5)$$

$$S'(\chi) = - ((1-\sin\chi)/2)\ln((1-\sin\chi)/2) - ((1+\sin\chi)/2)\ln((1+\sin\chi)/2)$$

$$\chi = \arccos(\exp(-(\pi/2)\theta_1)), \arccos(\exp(-(\pi/2)\theta_2))$$

The EPS recoherence is

$$\Delta S = S(\theta_1) - S(\theta_2) \qquad (6)$$

The recoherence is a positive quantity, because the entropy S is[12] monotonically increasing.
The environment decoherence is

$$\Delta S' = S'(\theta_1) - S'(\theta_2) \qquad (7)$$

The decoherence is a negative quantity, because the entropy S' is[12] monotonically decreasing.
The decoherence and recoherence are related by the inverse temperature $\beta$

$$\beta = - \Delta S' / \Delta S \qquad (8)$$

Viewing the decoherence and recoherence as a forward and a reverse process[13] gives (8).
Let the quantities $P_1^{am}$, $P_2^{am}$, $P_1^{pm}$, $P_2^{pm}$ be set as follows

$$P_1^{am} = \min \{ \max_\tau (P^a_t(\tau)), \max_\tau (P^b_t(\tau)), \max_\tau (P^c_t(\tau)) : \tau = 1,\ldots,12 \} \qquad (9)$$

$$P_2^{am} = \max \{ \max_\tau (P^a_t(\tau)), \max_\tau (P^b_t(\tau)), \max_\tau (P^c_t(\tau)) : \tau = 1,\ldots,12 \}$$

$$P_1^{pm} = \min \{ \max_\tau (P^a_t(\tau)), \max_\tau (P^b_t(\tau)), \max_\tau (P^c_t(\tau)) : \tau = 13,\ldots,24 \}$$

$$P_2^{pm} = \max \{ \max_\tau (P^a_t(\tau)), \max_\tau (P^b_t(\tau)), \max_\tau (P^c_t(\tau)) : \tau = 13,\ldots,24 \}$$

The EPS work for a day is

$$W_1 = 11.608 + ( \ln(P_1^{pm}) - \ln(P_1^{am})) / \beta \qquad (10)$$

$$W_2 = 11.608 + ( \ln(P_2^{pm}) - \ln(P_2^{am})) / \beta$$

This work is obtained from the transient fluctuation relation.[14] Here $\beta$ is the inverse temperature from (8).

4. **Testing the Daily Thermodynamics**

The time test of the daily EPS thermodynamics are the maximum likelihood seasonal cointegration tests for daily data.[15]

Let $T_{6,1}$, $T_{6,2}$, $T_{16}$, $T_{24}$ be the following times

$$T_{6,1} = 2^i W_1(\pi/2)\theta_1\tanh((\pi/2)\theta_1) \qquad (11)$$

$$T_{6,2} = 2^k W_2(\pi/2)\theta_2\tanh((\pi/2)\theta_2)$$

$$T_{16} = 2^m W_2(\pi/2)\theta_2\tan((\pi/2)\theta_2)$$

$$T_{24} = 1.5^n W_1(\pi/2)\theta_1\tan((\pi/2)\theta_1)$$

In (11), i is an integer, such that $T_{6,1} \in (0,9)$, k is an integer, such that $T_{6,2} \in (0,9)$, m is an integer, such that $T_{16} \in (10, 20)$, and n is an integer, such that $T_{24} \in (20, 30)$. In (11), $W_1$ and $W_2$ are from (10), and $\theta_1$ and $\theta_2$ are from (4).

The times $T_6$, $T_{16}$ and $T_{24}$ are obtained for the EPS evolution. They are set by the parity violation under evolution.[16] Here the EPS evolution follows a spiral, equivalent to the spiral obtained by Imel'baev and Chernysh[17] under coarsening of a system with loops.

The time $2T_6$ is a maximum likelihood statistic for cointegration. Its critical value is the critical value of Darné,[15] at 5% acceptance region, at the first level of seasonal cointegration, at 260 sample size and under a basic regression model with a constant, seasonal dummies and no trend. Then, the critical value of the time $2T_6$ is that value, from among the values 8.11, 11.10 and 11.30, compared to which the time $2T_6$ is smaller.

The time $T_{16} \geq 16$ is a maximum likelihood statistic for cointegration. Its critical value is the critical value of Darné,[15] at 5% acceptance region, at the second level of seasonal cointegration, at 260 sample size and under a basic regression model with a constant, seasonal dummies and no trend. Then, the critical value of the time $T_{16}$ is that value, from among the values 15.11, 18.01 и 18.18, compared to which the time $T_{16}$ is smaller.

The time $T_{16} < 16$ is a maximum likelihood statistic for cointegration. Its critical value is the critical value of Darné,[15] at 10% acceptance region, at the first level of seasonal cointegration, at 260 sample size and under a basic regression model with a constant, seasonal dummies and no trend. Then, the critical value of the time $T_{16}$ is that value, from among the values 11.95, 15.05 и 15.36, compared to which the time $T_{16}$ is smaller.

The time $T_{24} \geq 24$ is a maximum likelihood statistic for cointegration. Its critical value is the critical value of Darné,[15] at 5% acceptance region, at the third level of seasonal cointegration, at 260 sample size and under a basic regression model with a constant, seasonal dummies and no trend. Then, the critical value of the time $T_{24}$ is that value, from among the values 21.82, 24.64 и 24.73, compared to which the time $T_{24}$ is smaller.

The time $T_{24} < 24$ is a maximum likelihood statistic for cointegration. Its critical value is the critical value of Darné,[15] at 10% acceptance region, at the second level of seasonal cointegration, at 260 sample size and under a basic regression model with a constant, seasonal dummies and no trend. Then, the critical value of the time $T_{24}$ is that value, from among the values 19.40, 22.64 и 22.77, compared to which the time $T_{24}$ is smaller.

The time test of the daily EPS thermodynamics consists in a critical value check of each of the times $T_6, T_{16}, T_{24}$.

The energy test of the daily EPS thermodynamics is a test for an energy reserve R

$$R_1 = \exp(W_{0,1}\beta) - (2/(1 + W_{0,1}^{1/2}))^{1/2} \tag{12}$$

$$R_2 = \exp(W_{0,2}\beta) - (2/(1 + W_{0,2}^{1/2}))^{1/2}$$

$W_{0,1} = W_1 - 11.608$, $W_{0,2} = W_2 - 11.608$

This test follows from the hypergeometric function inequalities,[18] from the presentation of energy as a hypergeometric function[19] and from the time-independent relations in non-equilibrium systems.[20]

The energy test of the daily EPS thermodynamics consists in checking the positiveness of the reserve R.

5. **Thermal Computation**

The evolution behaviour can be presented as a statistical submanifold evolution surface, using the reversible entropic dynamics.[21] The mean and the standard deviation of the EPS evolution behaviour, by Cafaro and al.,[21] are

$$\mu = (1/W_1^{1/2})(\cosh((\pi/2)\theta_1) - \sinh((\pi/2)\theta_1)) \tag{13}$$

$$\sigma = (1/W_2^{1/2})(\cosh((\pi/2)\theta_2) - \sinh((\pi/2)\theta_2))/(\cosh(\pi\theta_2) - \sinh(\pi\theta_2) + 1/(8W_2((\pi/2)\theta_2)^2))$$

Here $\theta_1, \theta_2$ are angles from (4), and $W_1, W_2$ is work from (10).

The diffusion of the EPS evolution behaviour gives the following expected daily prices of electricity

$$c_1 = 10\beta\sigma_1 \tag{14}$$

$$c_2 = 10R_1\sigma_1 \tag{15}$$

$$c_3 = 10R_2\sigma_1 \tag{16}$$

In (14), $\beta$ is the inverse temperature from (8). In (15),(16), $R_1$, $R_2$ is the reserve from (12). In (14),(15),(16), $\sigma_1 = \sigma$, if $\sigma > 1$, and $\sigma_1 = 2-\sigma$, if $\sigma < 1$. Here $\sigma$ is from (13).

These daily prices of electricity have been found as prices on a market in uncertainty by Pennock and al.[22] The multiplication by ten in (14) is photographic enlargement, made by Grenander,[23] of the price on the market in uncertainty to a price on the electricity market for a day ahead.

These three prices set the following prices:

$$c_a = 2c_1/3 \tag{17}$$

$$c_m = \min(c_2, c_3)$$

$$c_s = \max(c_2, c_3)$$

The expected EPS reliability, with respect to a rare event and a competitive phenomenon, is

$$p_r = c_a / c_s, \text{ if } c_a < c_s \tag{18}$$

$$p_r = c_s / c_a, \text{ if } c_a > c_s$$

This reliability is found as a Jordan curve descriptor introduced by Zuliani and al.[24]

The expected EPS reliability, with respect to a cooperative phenomenon, is

$$p_v = 2(1 - p_w) \tag{19}$$

$$p_w = 1 - 0.5(c_m / c_a)^{\frac{1}{2}}, \text{ if } c_a < c_s$$

$$p_w = 0.5(c_a / c_m)^{\frac{1}{2}}, \text{ if } c_a > c_s$$

This reliability is found as a non-stationary realization descriptor introduced by Daneev and al.[25]

The expected EPS droop is

$$k_c = 1.261060863 \, \beta \tag{20}$$

Here $k_c$ is set by the inverse temperature $\beta$ from (8) for the EPS scheme, viewed as a Euclidean 4-design by Bannai and Bannai.[26]

The expected daily price of electricity, with respect to the EPS reliability, is

$$c_H = 1000 p_w / (50 p_r - 2\pi k_c) \tag{21}$$

This daily price of electricity minimizes the EPS lifetime variance, in accordance with Ref.27.

The computation by thermalisation reduces the daily mean error, relative to the daily peak load, of the load forecast, by $\delta$

$$\delta = 10(W_1 \mu / 1.78617 - \Delta S) \tag{22}$$

Here $\delta$ is in %, $W_1$ is from (10), $\mu$ is from (13), $\Delta S$ is from (6), and the normalization of $\mu$ is from the Euclidean 4-designs by Bannai and Bannai.[26] The quantity $\delta$ is determined by the computational potential introduced by Anders and al.[28] The multiplication by ten in (22) is photographic enlargement, made by Grenander[23], of the computational potential to an EPS potential.

## 6. Daily Artificial Dispatcher

The "Daily Artificial Dispatcher" (DAD) is a field intelligent system which makes the thermal predictive analysis set out above. This thermal predictive analysis is a predictive analysis of the EPS security because it gives the expected EPS load from (1), (2) and (3), the expected electricity price from (14) and (21), the expected reserve from (12), the expected droop by (20) and the expected EPS reliability from (18), (19).

Predicting the times (11) is predicting the synchronization in the EPS. Predicting the energy reserve (12) is predicting the stability in the EPS. DAD's predicting the synchronization in the EPS shows that DAD perceives the cointegration. DAD's predicting the stability in the EPS shows that DAD interprets the cointegration. DAD is then an intelligent system in the sense of Ref.29. This intelligence is wisdom because it consists in evasion and prediction. Indeed, falling out of synchronization is evasion, and stability is prediction.

The thermalisation in finding the EPS security is a field computation according to Ref.28 and Ref.4. Therefore, DAD is indeed a field intelligent system.

DAD's resource is heat. That is why DAD's logic is the logic of resources, i.e. Girard's linear logic. This conclusion is natural because of the connection[30] between evasion/prediction and linear logic.

DAD has the wisdom of electricity consumers. DAD presents the average belief of these consumers about the evolution in the EPS behaviour based on an expected warming of the weather. The consumers' average belief is that the EPS reliability, with respect to a cooperative phenomenon, is $p_v$ from (19). The consumers' average belief is that the EPS reliability, with respect to a rare event and a competitive phenomenon, is $p_r$ from (18).

DAD stakes the following part of its resources on the EPS reliability $p_r$ with respect to a rare event and a competitive phenomenon

$$s_r = p_v - p_r p_v^* / p_r^*, \text{ if } p_v p_r^* > p_r p_v^* \qquad (23)$$

$$s_r = 0, \text{ if } p_v p_r^* \leq p_r p_v^*$$

DAD stakes the following part of its resources on the EPS reliability $p_v$ with respect to a cooperative phenomenon

$$s_v = 0, \text{ if } p_v p_r^* > p_r p_v^* \qquad (24)$$

$$s_v = p_r - p_v p_r^* / p_v^*, \text{ if } p_v p_r^* \leq p_r p_v^*$$

Thus DAD acts[31] as a rational forecast gambler.

In (23),(24), $p_r^*$ and $p_v^*$ are set by the sufficient conditions given by Wolfers and Zitzewitz,[32] under which prediction market prices coincide with the average beliefs among traders.

The quantity $p_r^*$ is obtained from the following equation

$$(\sigma - 1)x^3 + (2 - \sigma)x^2 - \sigma x + 2(\sigma - 1)p_v = 0 \qquad (25)$$

In (25), σ is the standard deviation from (13).

This quantity is set by the greatest positive root $p_r^o$ of the equation (25)

$$p_r^* = p_r^o, \text{ if } p_r^o \leq 1 \qquad (26)$$

$$p_r^* = 2 - p_r^o, \text{ if } p_r^o > 1$$

The quantity $p_v^*$ is obtained from the following equation

$$(\sigma - 1)x^3 + (2 - \sigma)x^2 - \sigma x + (\sigma - 1)p_r = 0 \qquad (27)$$

In (27), σ is the standard deviation from (13).

This quantity is set by the greatest positive root $p_v^o$ of the equation (27)

$$p_v^* = 1 - p_v^o / 2, \text{ if } p_v^o \leq 1 \qquad (28)$$

$$p_v^* = p_v^o / 2, \text{ if } p_v^o > 1$$

DAD checks the expected EPS synchronization by the time test of the daily EPS thermodynamics. DAD checks the expected EPS stability by the energy test of the daily EPS thermodynamics. Thus DAD verifies its forecasts.

## 7. The Results of DAD

DAD operates to help the dispatchers of the Bulgarian EPS.

The regressions (1), (2) and (3) are estimated from the hourly sampled values of the daily load and the dry bulb temperature for the preceding nine days, as well as from an hourly sampled daily forecast of the dry bulb temperature.To estimate these regressions, use is made of the load data, supplied by the Bulgarian electric power system operator ESO EAD and AccuWeather's weather forecast, used by ESO EAD. An estimate of the three regressions is obtained by the exact maximum likelihood method for dynamic regression estimation given by Pesaran and Slater.[33] This estimate is correct because the sample is non-stationary and of small size.

The mean error, relative to the daily peak load, of the daily load forecast of the Bulgarian EPS, obtained by dynamic regression, is 5%. DAD reduces this error by 1.5% down to 3.5%.

Table 1 gives the monthly average daily mean error, relative to the daily peak load, of the Bulgarian EPS load forecast made by DAD. This error is given for two months of the year, choosing those months where the error is maximal.

Table 1. The monthly average daily mean error, relative to the daily peak load, of the Bulgarian EPS load forecast made by DAD, in percent.

| Year  | 2000 |      | 2001 |      | 2002 |      | 2003 |      | 2004 |      | 2005 |      | 2006 |      |
|-------|------|------|------|------|------|------|------|------|------|------|------|------|------|------|
| Month | 04   | 10   | 04   | 10   | 04   | 10   | 04   | 09   | 05   | 10   | 04   | 10   | 04   | 10   |
| MMRE  | 3.07 | 3.24 | 3.90 | 2.62 | 3.23 | 3.33 | 3.83 | 2.90 | 3.14 | 3.09 | 2.83 | 2.86 | 3.39 | 2.87 |

DAD reduces the mean error, relative to the daily peak load, of the Bulgarian EPS daily load forecast by 1.5%. This reduction corresponds to a hypothetic increase in the average daily temperature by $0.7°C$. Here it is assumed that a $2°C$ increase in the average daily temperature leads to a 4.6% change, in accordance with the results of Crowley and Joutz.[34]

A hypothetic warming of the weather by $0.7°C$ for a day ahead results in a correct prediction of the expected EPS security by DAD. For example, the expected zero reliability for a cooperative phenomenon gives a true prediction for an EPS decoupling.

## 8. Conclusion

The aim of this paper is a thermal predictive analysis of the EPS security for a day ahead. This aim has been achieved as follows:

1/ one descriptive and two rescriptive dynamic models for prediction of the daily load have been constructed. These models have been obtained by cointegration of the daily EPS load and the weather;

2/ the daily EPS thermodynamics has been found through the EPS inverse temperature and through the EPS work;

3/ a time test and an energy test of the daily EPS thermodynamics have been obtained;

4/ thermal computation of the expected EPS security for a day ahead has been made;

5/ the predictive analysis of the EPS security for a day ahead has been presented as a field intelligent system that shows to the EPS dispatchers what the electricity consumers' wisdom is;

6/ it has been shown that the proposed predictive analysis enhances the EPS security by more accurate prediction of the EPS load and by the prediction of critical phenomena.